\documentclass[aps,prl,twocolumn,floatfix,superscriptaddress]{revtex4}
\usepackage{graphicx}
\usepackage{epsfig}
\begin{document}
\newcommand{\norm}[1]{\ensuremath{| #1 |}}
\newcommand{\aver}[1]{\ensuremath{\langle #1 \rangle}}
\newcommand{\ket}[1]{\ensuremath{| #1 \rangle}}
\title{Spin-charge separation in two-component Bose-gases}

\newcommand{\aachen}{Institute for Theoretical Physics C, RWTH Aachen,
  D-52056 Aachen, Germany}
\newcommand{\geneve}{DPMC-MaNEP, University of Geneva, 24 Quai
  Ernest-Ansermet, CH-1211 Geneva, Switzerland}

\author{A. Kleine} \affiliation{\aachen}
\author{C. Kollath} \affiliation{\geneve}
\author{I.P. McCulloch} \affiliation{\aachen}
\author{T. Giamarchi} \affiliation{\geneve}
\author{U. Schollw\"ock} \affiliation{\aachen}

\date{\today}  

\begin{abstract}

We show that one of the key characteristics of interacting one-dimensional
electronic quantum systems, the separation of spin and charge, can be observed
in a two-component system of bosonic ultracold atoms even close to a competing phase
separation regime. To this purpose we determine the real-time evolution of a
single particle excitation and the single-particle spectral function using
density-matrix renormalization group techniques. Due to efficient bosonic
cooling and good tunability this setup exhibits very good conditions for observing this
strong correlation effect. In anticipation of experimental
realizations we calculate the velocities for spin and
charge perturbations for a wide range of parameters.

\end{abstract}
\maketitle
%%%%%%%%%%%%%%%%%%%%%%%%%%%%%%%%%%%%%%%%%%%%
One of the most exciting recent events is the ever-growing
interplay
between previously disconnected fields of physics, such as quantum optics
and condensed matter physics. In particular, cold
atomic systems have opened the way to engineer strongly interacting quantum
many-body systems of unique purity. The unprecedented control over interaction strength
and dimensionality allows the realization of ``quantum simulators'' where
fundamental
but hard to analyze phenomena in strongly correlated systems could be
observed and
controlled. Examples are the
observation
of superfluid to Mott insulator transition for Bose gases
\cite{GreinerBloch2002} and the fermionization
of strongly interacting one dimensional bosons
\cite{ParedesBloch2004,KinoshitaWeiss2004}.

Among interacting systems, the physics depends very strongly on the dimensionality. In one-dimensional systems the interactions play a major role and lead to drastically different
physics than
for their higher dimensional counterpart. Typically, interactions
in one-dimensional systems lead to a Luttinger liquid state where the excitations of the system are
collective excitations \cite{Giamarchibook}.
The importance of such a state for a large variety of experimental devices
in condensed matter has
led to a hunt to observe its properties. A remarkable consequence of such a state
is the absence of single particle excitations. This means that a quantum
particle, that would normally carry both
charge and spin degrees of freedom, fractionalizes into two different
collective excitations,
a spin and a charge excitation. Such a fractionalization of a single
particle excitation is the hallmark of collective effects caused by
interactions. However, just as detecting fractional excitations in the case of the quantum hall
effect is difficult \cite{glattli_fqhe_review}, 
observing spin-charge separation has proven elusive despite several
experimental attempts
\cite{Segovia1999,Lore02,kim_spincharge_photoemission}. 
So far, the best experimental evidence is provided by tunneling between
quantum wires where interferences effects 
are due to the existence of two different velocities \cite{Ausl05}. However,
in these systems it is hard to quantify
or to tune the interaction between the particles which causes the collective
effects. Since control of the interaction is a routing procedure in ultracold gases, the
possible realization
of the phenomenon of spin charge separation has also been discussed in the context of cold fermionic
gases \cite{RecatiZoller2003,KeckeHaeusler2004,KollathZwerger2005,KollathSchollwoeck2006b} and strongly interacting bosonic gases \cite{ParedesCirac2003}.

However, proposals to observe spin-charge separation in ultracold fermionic
gases are still plagued by the currently quite high temperatures in
such systems. A much better setup to test for spin-charge separation would be 
two-component Bose gases, for example using the $\ket{F=2,m_F=-1}$ and the
$\ket{F=1,m_F=1}$ hyperfine states of $^{87}$Rb \cite{ErhardSengstock2004,WideraBloch2004}.
Experimentally, this system retains the advantages of the fermionic
ultracold
atom setup while allowing for much lower temperatures due to the more
efficient cooling techniques available for bosons. However, theoretical
studies
\cite{CazalillaHo2003, MishraDas2006} for one-dimensional systems 
predict that close to the experimentally accessible parameter regime of almost equal
inter- and intra-species interaction strength phase
separation occurs. This is the remaining potential
experimental complication in the setup.

In this work we demonstrate the phenomenon of spin-charge separation in the
experimentally relevant parameter
regime, allowing to use this system to unambiguously test for spin-charge
separation. We calculate both the real-time evolution
of a single particle excitation and the
dynamical single particle spectral function of the two-component bosonic
systems. We show that both these quantities demonstrate the separation of a
single-particle excitation into spin and charge. We further
determine the velocity of spin and charge and the Luttinger parameters for
experimentally relevant parameter regimes. To perform the calculations we use variants of the density matrix renormalilzation group method
(DMRG) \cite{White1992, Schollwoeck2005}. The numerical treatment is necessary
to obtain reliable predictions for experimental realizations, due to the close
proximity of this regime to phase separation.

%%%%%%%%%%%%%%%%%%%%%%%%%%%%%%%%%%%%%%%%%%%%%%%%%%%%

A one-dimensional two-component Bose gas in an optical
lattice \cite{JakschZoller1998} can be described by the two-component Bose-Hubbard model
\begin{equation}
\label{eq:H}
\begin{array}{rl}
H =& -J \sum_{j,\nu} \left( b^\dagger_{j+1,\nu}b_{j,\nu} +
h.c.\right) + \sum_{j,\nu} \frac{U_\nu \hat{n}_{j,\nu}(\hat{n}_{j,\nu}-1)}{2}  \\
&+U_{12} \sum_j \hat{n}_{j,1}\hat{n}_{j,2}+ \sum_{j,\nu}
\varepsilon_{j,\nu} \hat{n}_{j,\nu} 
\end{array}
\end{equation} 
Here $j$ is the site index and $\nu= 1, 2$ labels the two different hyperfine states
of the system, $b$ and $b^\dagger$ are the annihilation and creation operators
and $\hat{n}$ is the number operator. The first term models the kinetic energy of
the atoms. The intra-species interaction is described by the
$U_\nu$ term. We use $U:=U_1=U_2$ as it is approximately
realized for commonly used hyperfine states of $^{87}$Rb \cite{WideraBloch2006}. The inter-species interaction is given by the $U_{12}$ term
and the last term describes external potentials. In the following we use
the dimensionless parameters $u=U/J$ and $u_{12}=U_{12}/J$. We define the
`charge' density $n_{j,c}=n_{j,1} +n_{j,2}$ and the `spin' density $n_{j,s}=n_{j,1} -n_{j,2}$. We focus on systems with average filling $n=\sum_j n_{j,1}/L=\sum_j n_{j,2}/L$ smaller
than one particle per site and parameter regimes up to close to the transition to the
phase separation (approximately $u_{12} \approx u$ \cite{MishraDas2006}). Here
$L$ is the number of sites in the system.

%%%%%%%%%%%%%%%%%bosonization%%%%%%%%%%%%%%%%%%%%
In a superfluid phase away from the transition to phase separation the low
 energy physics can be approximated by a density-phase representation of the
 bosons as used in the bosonization method \cite{Giamarchibook}. In this representation the Hamiltonian is totally separated into one part for the
charge and one for the spin degrees of freedom. The physics is fully determined
by the velocities $v_{c,s}$ and the so-called Luttinger
parameters $K_{c,s}$ for spin (s) and charge (c). Therefore the separation of
a single particle excitation into spin and charge excitations is expected. 
The parameters of two interacting species of bosons can be related to the
parameters $K$ and $v_0$ for the single species case
\cite{OrignacGiamarchi1998} by
\begin{eqnarray}
\label{eq:vel}
v_{c,s}&&=v_{0}\sqrt{1\pm (g_{12}K)/(\pi v_0)}\\ 
\textrm{and } K_{c,s}&&=K / \sqrt{1\pm
  (g_{12}K)/(\pi v_0)}. \nonumber
\end{eqnarray}
Here $g_{12}$
is the interspecies interaction strength in the continous model.  
In the limit of small interactions, the single species parameters $K$ and
$v_0$ can be directly related to the Bose-Hubbard Hamiltonian Eq.~(\ref{eq:H})
\cite{Giamarchibook}. For higher values of the interaction strength the relation even for the single
species situation is unkown, and has to be determined numerically \cite{KollathZwerger2004}. For large values of the inter-species interaction
of the order of the intra-species interaction the system approaches the transition to phase
separation and the bosonization approach becomes a priori inaccurate. 

\begin{figure*} 
\begin{center}
         {\epsfig{figure=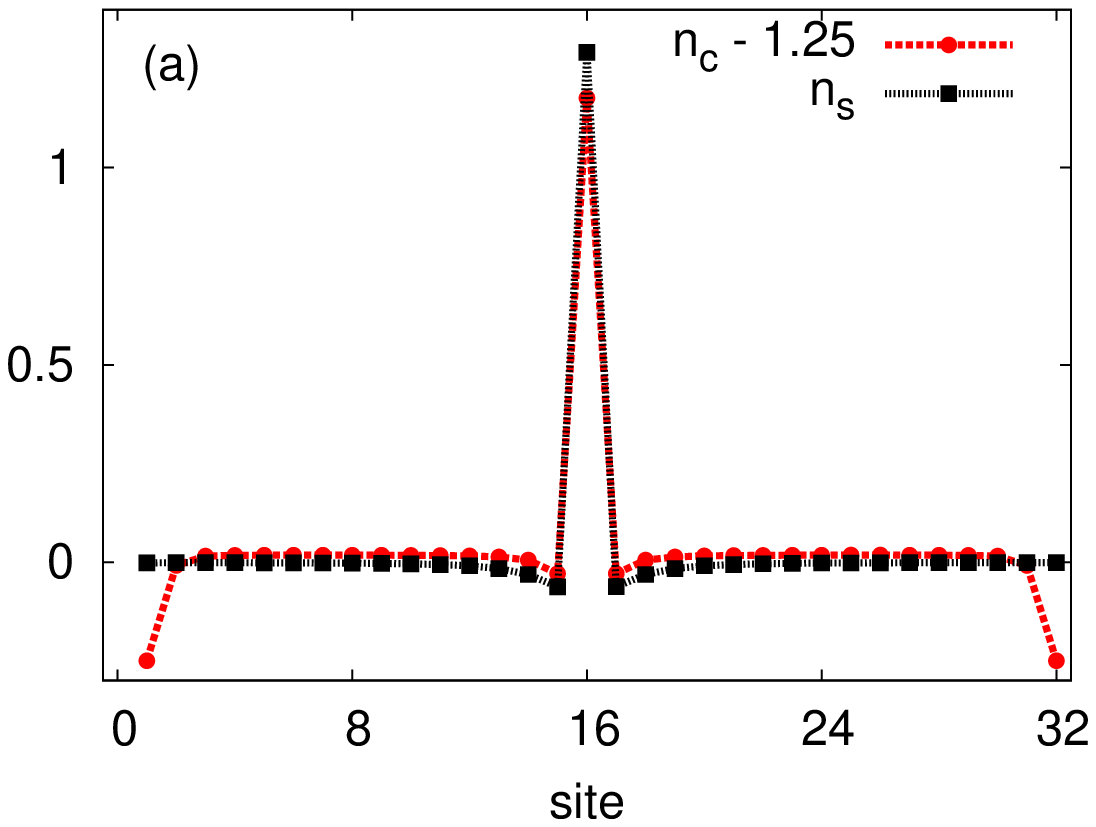,width=0.3\linewidth}}
	 {\epsfig{figure=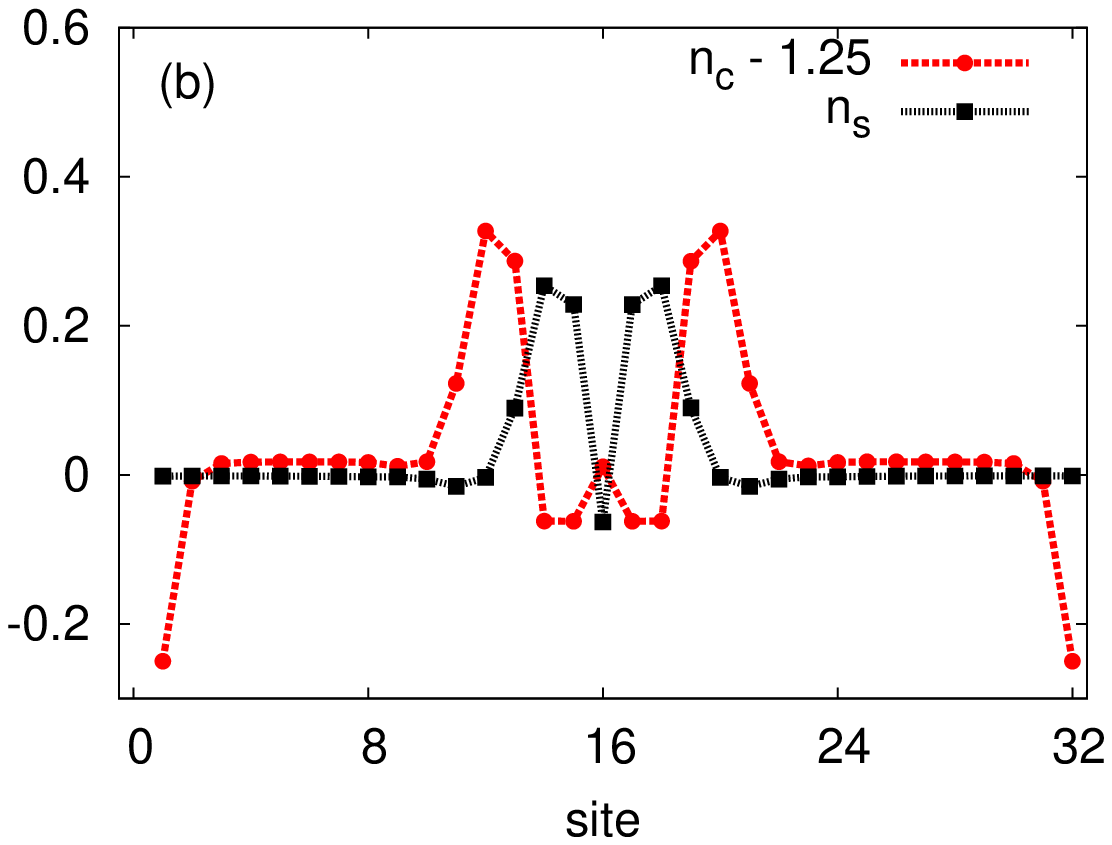,width=0.3\linewidth}}
	 {\epsfig{figure=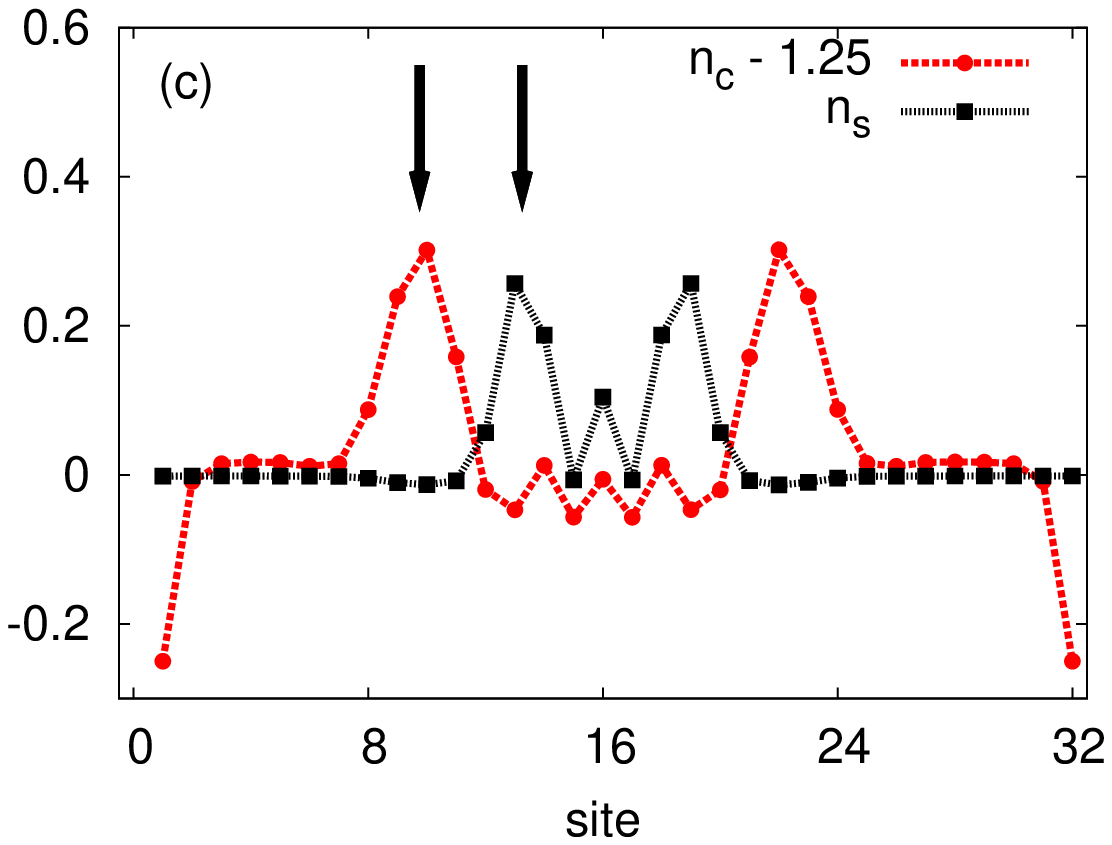,width=0.3\linewidth}}
\end{center}
\caption{(Color online) Snapshots of the time-evolution of the charge and spin density
  distribution of a single particle excitation created at time $t=0\hbar/J$; (a)
  at time $t=0\hbar/J$, (b) at time $t=1.5 \hbar/J$ and (c) at time $t=2.5
  \hbar/J$. The system parameters were $n_{1,2} = 0.625$, $u = 3$, $u_{12} = 2.1$.
  The charge density is shifted by $1.25$ for better
  visibility. The arrows in (c) mark the clear separation of the charge and the spin
  density waves. }
\label{fig:single}
\end{figure*}

%%%%%%%%%%%%%%single particle excitation%%%%%%%%%%%%
%%%%%%%%%%%%real time evolution%%%%%%%%%%%%%%%%%%
Snapshots of the real time-evolution of a single particle excitation in a two-component
bosonic system are shown in Fig.~\ref{fig:single}. The single particle excitation at time $t=0$ is prepared by the
application of the creation operator of one species, say $1$, on site $L/2$ to the
ground state $|\psi_0\rangle$, i.e.~$b_{L/2,1}^\dagger|\psi_0\rangle$. The
resulting sharp peaks in the density distributions are shown in
Fig.~\ref{fig:single} (a). For $t>0$ the time-evolution of the single particle excitation is
calculated using the adaptive time-dependent DMRG
\cite{DaleyVidal2004,WhiteFeiguin2004}. The time-evolution is performed using
a Krylov algorithm \cite{HochbruckLubich1997} in a matrix product state basis
with a fixed error bound for each
timestep of the order of $\left\| \ket{\Psi(t+\Delta t)} - \exp[-i H \Delta
  t] \ket{ \Psi(t)} \right\|^2<10^{-5}$. The stepsize
$\Delta t = 0.2$ and $6$ to $10$ Krylov vectors were used resulting in
 Hilbert spaces with a local dimension of a few thousand states. As can be seen in the
 snapshots in Fig.~\ref{fig:single} the initial single particle excitation
splits up into two counter propagating density waves. Due to their different
spin and charge velocities, after a period of time a clear
separation of the density waves is seen
(cf.~Fig.~\ref{fig:single} (c)) \footnote{The remaining interaction between the spin
and the charge degrees of freedom for short times can cause small additional
structures beside the main peaks in the density wave profiles [Fig.~\ref{fig:single} (c)].}.

\begin{figure} 
\begin{center}
        {\epsfig{figure=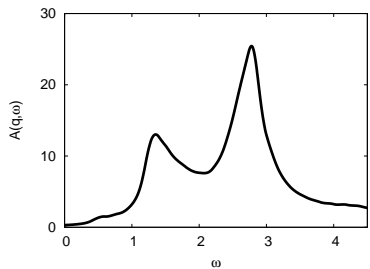,width=0.6\linewidth}}
\end{center}
\caption{One-particle spectral function at momentum $q = 20/65 \pi/a$. Two peaks corresponding to the spin
  and the charge excitation can be distinguished. The following parameters
  were used $n = 0.625$, $u = 3$, $u_{12} = 2.1$ on a system with
  $L=64$ sites and a broadening $\eta=0.1$. Here $a$ is the lattice spacing.}
\label{fig:spectral}
\end{figure}

%%%%%%%%%%%%%%%%spectral function%%%%%%%%%%%%%%%%%%%%
Additionally to the time-evolution of a single particle excitation, we
obtained the single particle spectral function
$A_\nu(q,\omega)=\frac{1}{\pi} \Im \aver{b_{q,\nu} ^\dagger
  \frac{1}{H+\omega+\i\eta-E_0}b_{q,\nu} } $ as shown in
Fig.~\ref{fig:spectral}. For fermions this function is known to exhibit two peaks at the spin and
charge excitation energies
\cite{MedenSchoenhammer1992,Voit1993},
showing a direct signature of the spin-charge separation. 
For the bosons computing this spectral function is more involved and up to
very recently it was only derived for a single component bosonic system 
\cite{haldane_bosons,KuehnerWhite1999,CauxCalabrese2006}. An expression for
the correlation
functions which allow to obtain the single-particle correlation function 
within the bosonization treatment for a two-component bosonic system was
derived in \cite{IucciGiamarchi2007}. 
Power law singularities at $q v_{c,s}$ are obtained with respective exponents
$1/4K_{c,s}+1/2K_{s,c} - 1$. For the values of the Luttinger
parameters (as shown in Fig.~\ref{fig:spectral}) one thus expects two divergent peaks. We show in Fig.~\ref{fig:spectral} the full spectral function for our
microscopic model, as calculated numerically using a matrix product state generalization of
the correction vector method\cite{KuehnerWhite1999,FriedrichSchollwoeck2007}.
Our results show clearly the
appearance of the two separated peaks, the lower representing the spin and
the upper the charge excitation branch \footnote{Given the relatively large
  value of $q$ that we use and the band curvature that exists in the microscopic model it is difficult 
to quantitatively compare the position of the peaks to the bosonization 
result $u_{c,s} q$, valid for small $q$ or a strictly linear dispersion 
relation.}. Thus both the real time evolution of a single particle function and the
single particle spectral function show clear signatures of separation of
spin and charge.

%%%%%%%%%%%%%velocities%%%%%%%%%%%%%%%%%%%
To observe the separation of spin and charge excitations experimentally in a
system of ultracold bosons, knowledge of the spin and charge velocities is
indispensable. We therefore determined the velocities for a wide range of parameters. This was done
calculating the time-evolution of a small spin and charge density perturbation
respectively. The density perturbation was created at time $t=0$ applying
an external potential of Gaussian form
$\varepsilon_{\nu,j}= \varepsilon_0 \exp \left( - \frac{\left(j - j_0\right)^2}{2\sigma_j}\right) $
where $\nu = c, s $. At time $t=0$ the potential
is switched off and the time-evolution of the density perturbation is
calculated. The errors in the obtained velocities are of the order of $0.01 aJ/\hbar$
for small $u_{12}$ and increase with larger $u_{12}$.

\begin{figure} 
\begin{center}
         {\epsfig{figure=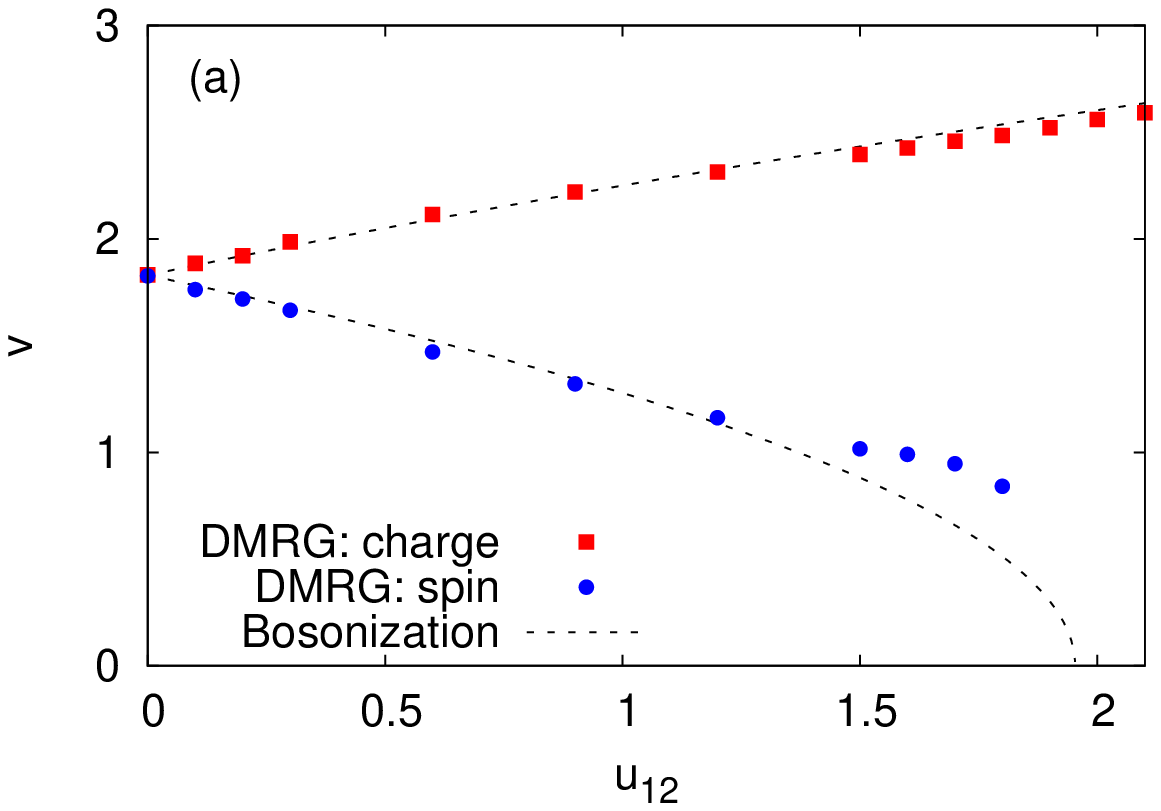,width=0.75\linewidth}}
{\epsfig{figure=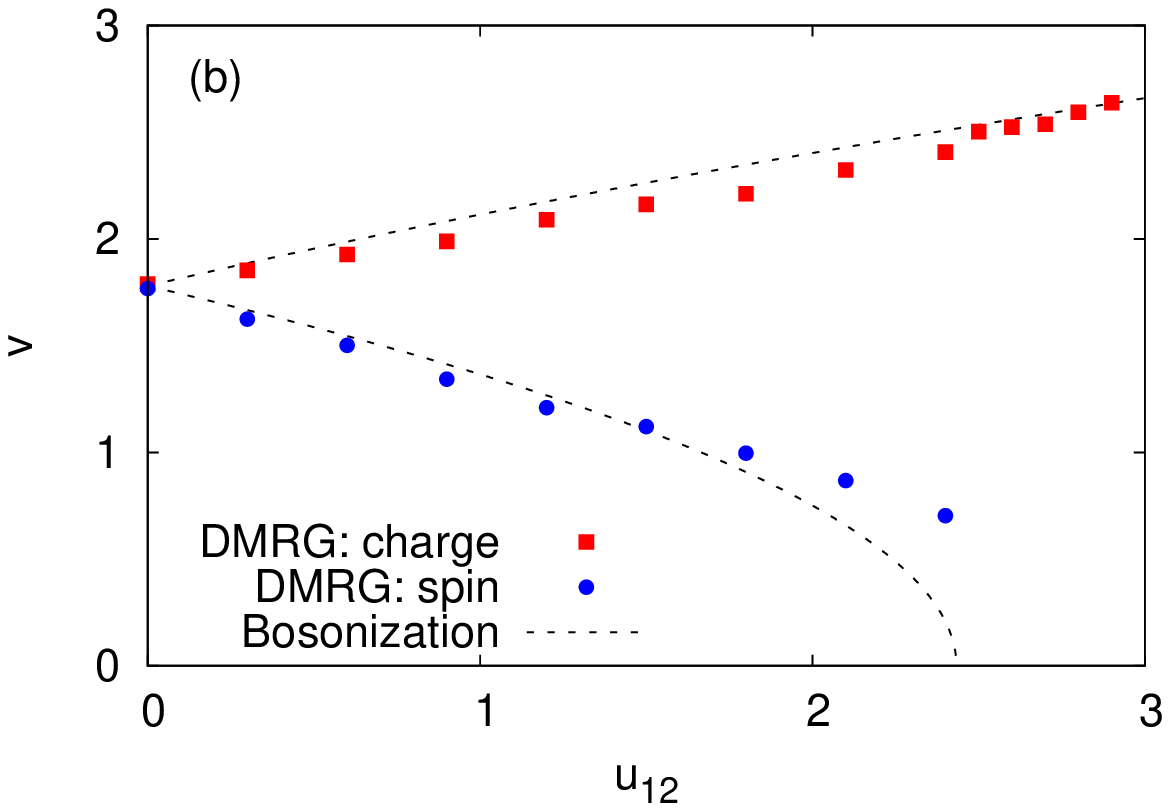,width=0.75\linewidth}}
\end{center}
\caption{(Color online) Dependence of the charge and spin velocity on the interparticle
  interaction strength for (a) $u_{12}=2$ and $n\approx 0.88$ and (b)
  $u_{12}=3$ and $n=0.63$.  A comparison of analytical results (line, see text) and
  numerical DMRG results (symbol) is shown. The velocities are measured in
  units $aJ/\hbar$. Note that the errors of the DMRG
  results increase close to $u_{12}\approx u$.}
\label{fig:vel}
\end{figure}

In Fig.~\ref{fig:vel} we show the dependence of the velocities on the
inter-species interaction for two different values of the parameter $\gamma=u/(2n_{c})$. For both parameter regimes, the
charge velocity increases with increasing interaction whereas the spin
velocity descreases. For a
vanishing inter-species interaction it was shown in \cite{KollathZwerger2004}
that for $\gamma< 1$ the bosonization and the solution of the
exactly solvable Lieb-Liniger model agree approximately with the velocities in
the Bose-Hubbard model. In Fig.~\ref{fig:vel} (a) we find for $\gamma\approx
1.1$ very good agreement with the
analytical solutions even up to strong inter-species interaction
strength $u_{12}$ close to phase separation. For $\gamma\approx 2.4$ [cf.~Fig.~\ref{fig:vel} (b)] even for vanishing interspecies
interaction the deviations from the direct relation between the parameters
in the Bose-Hubbard model are considerable (cf.~\cite{KollathZwerger2004}). However, we find that the
dependence of the velocities on the interspecies interaction strength via Eq.(~\ref{eq:vel}) is still
a good approximation correcting by the numerically determined value for
$v_0$ and $K$. This holds even up to close to the regime of phase separation, i.e.~$u_{12}\approx u$ where the
difference in the velocities is maximal. However, for $u_{12}\approx u$
the results for the spin velocity start to deviate for both values of $\gamma$. 

\begin{figure} 
\begin{center}
{\epsfig{figure=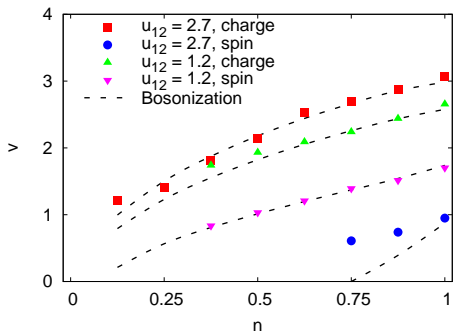,width=0.75\linewidth}}\\
\end{center}
\caption{(Color online) Dependence of the charge and spin velocity on the charge background density., comparison analytical results of
  bosonization and numerical DMRG results. The parameters used are (a) $u=3$, $u_{12}=1.2$ and (b) $u=3$, $u_{12}=2.7$. }
\label{fig:vel_n}
\end{figure}

In Fig.~\ref{fig:vel_n} we show for two different fixed inter-species
interaction strengths the
dependence of the velocities on the density.
The charge and the spin velocities rise with increasing background charge density. (Note, even
at $n_c=1$ the system is in the superfluid regime.) The increase of the
velocities is described to good accuracy using the analytical form
Eq.~(\ref{eq:vel}), provided we use numerically obtained
values of $K$ and $v_0$. For large $u_{12}$ and small $n_c$ the results for
the velocities from
DRMG, in particular the spin velocity, deviate considerably from Eq.~(\ref{eq:vel}), showing that the
approximate
relation cannot be used in this regime.
Note that in this regime the extraction of the spin velocity from the
real-time evolution becomes also more involved since the spin perturbation
shows
a strong spreading in space (cf.~\cite{PoliniVignale2007}). At the timescales
over which we calculated the velocity, the left- and the right-moving spin
perturbations are not yet fully separated and show strong amplitude damping.
Our finding of the dependencies can be
used to predict the velocities for experimentally interesting parameter
regimes.

%%%%%%%%%%%%%%%%%%%%%%Experiments%%%%%%%%%%%%%%%%%%

In recent experiments for the preparation of a mixture of two bosonic
components in optical lattices mostly two hyperfine states of $^{87}$Rb are
used, e.g.~the $\ket{F=2,m_F=-1}$ and the
$\ket{F=1,m_F=1}$ hyperfine states. The intra-species scattering lengths are
$a_2=91.28a_B$ and $a_1=100.4a_B$ \cite{WideraBloch2006}, respectively where $a_B$ is the Bohr radius. 
For these states the inter-species scattering length is of the
same order of magnitude as the intra-species scattering length and can be
tuned about 20\% using a Feshbach resonance \cite{WideraBloch2004,ErhardSengstock2004}. Thereby the
experimental parameters are close to the competing phase separation
regime. These mixtures can be confined to one-dimensional structures using
strongly anisotropic lattices
\cite{MoritzEsslinger2003,TolraPorto2004,KinoshitaWeiss2004,ParedesBloch2004}. The most intuitive observation of
the phenomenon of spin-charge separation in these systems is to generate a
single particle excitation and then follow the evolution of the excitation in
real time. This can be done measuring the spin-resolved density over a certain
region. The creation of a single particle excitation can be done 
 e.g.~using
outcoupling of single particles by the application of a magnetic field
gradient for addressability and a microwave field \cite{BlochEsslinger2000,OettlEsslinger2005,FoellingBloch2006}\footnote{The
magnetic field gradient can be applied since the two hyperfine states have
approximately the same magnetic moment.}. The efficiency of such a technique for
generating single particle exciations was demonstrated \cite{OettlEsslinger2005} using a cavity. The microwave field could be chosen to
couple the $\ket{F=1,m_F=1}$ hyperfine state to e.g.~the
$\ket{F=2,m_F=2}$. This has the advantage that scattering with the
$\ket{F=1,m_F=1}$ state are supressed. The
measurement of the density resolved over a region of approximately 10 lattice
sites can then be performed using again the magnetic field
gradient to get an unambigious signal. In an array of one-dimensional tubes the
broadening of the signal caused by the trapping potential 
could be suppressed by preparing most of the tubes in a Mott-insulating state as
shown in
\cite{KollathZwerger2005}. 

%%%%%%%%%%%%%%%%%%%%%%end Experiments%%%%%%%%%%%%%%%%%%

We would like to thank J.S.~Caux, S. F\"olling, M. K\"ohl, and B. Paredes for fruitful
discussions. AK and US acknowledge support by the DFG and CK and TG by the Swiss National Fund under MaNEP and
Division II and the CNRS. CK thanks the Institut
Henri Poincare for its hospitality during the final part of the work.

%\bibliography{references}

\end{document}